\documentclass[12pt]{article}
\usepackage{amssymb,amsmath}
\usepackage{epsfig}
\usepackage{url}
\usepackage[dvipdfmx]{hyperref}
\makeatletter
\def\alt{\mathrel{\mathpalette\gl@align<}}
\def\agt{\mathrel{\mathpalette\gl@align>}}
\def\gl@align#1#2{\lower.6ex\vbox{\baselineskip\z@skip\lineskip\z@
\ialign{$\m@th#1\hfil##\hfil$\crcr#2\crcr\sim\crcr}}}
\makeatother

\newcommand{\doi}[1]{\href{https://doi.org/#1}{doi:#1}}

\begin{document}

\begin{flushright}
January, 2020
\end{flushright}
\vspace*{2.0cm}
\begin{center}
\baselineskip 20pt
{\Large\bf
$\mu$\,-\,$\tau$ symmetry breaking and CP violation \\ in the neutrino mass matrix} \vspace{1cm}

{\large
Takeshi Fukuyama$^1$ and Yukihiro Mimura$^2$}
\vspace{.5cm}

$^1${\it
Research Center for Nuclear Physics (RCNP), \\ Osaka University, Ibaraki, Osaka, 567-0047, Japan
}\\
$^2${\it
Department of Physical Sciences, College of Science and Engineering, \\
Ritsumeikan University, Shiga 525-8577, Japan
}
\\

\vspace{.5cm}

\end{center}

\begin{center}{\bf Abstract}\end{center}

The $\mu$-$\tau$ exchange symmetry in the neutrino mass matrix and its breaking as a perturbation are discussed. 
The exact $\mu$-$\tau$ symmetry restricts the 2-3 and 1-3 neutrino mixing angles as
$\theta_{23} = \pi/4$ and $\theta_{13} = 0$ at a zeroth order level.
We claim that the $\mu$-$\tau$ symmetry breaking 
prefers a large CP violation to realize 
the observed value of $\theta_{13}$
and to keep $\theta_{23}$ nearly maximal,
though an artificial choice of the $\mu$-$\tau$ breaking can
tune $\theta_{23}$, irrespective of the CP phase. 
We exhibit several relations among the deviation of $\theta_{23}$ from $\pi/4$, 
$\theta_{13}$ and Dirac CP phase $\delta$,
which are useful to test the $\mu$-$\tau$ breaking models in the near future experiments.
We also propose a concrete model to break the $\mu$-$\tau$ exchange symmetry
spontaneously and its breaking is mediated by the gauge interactions radiatively
in the framework of the extended gauge model 
with $B-L$ and $L_\mu - L_\tau$ symmetries.
As a result of the gauge mediated $\mu$-$\tau$ breaking in the neutrino mass matrix,
the artificial choice is unlikely, and a large Dirac CP phase is preferable.

\thispagestyle{empty}

\bigskip
\newpage

\addtocounter{page}{-1}

\section{Introduction}
\baselineskip 18pt

The long baseline neutrino oscillation experiments are ongoing \cite{Abe:2018wpn,Acero:2019ksn}, and
it is expected that the 2-3 neutrino mixing and a CP phase
will be measured more accurately \cite{Abe:2018uyc,Abi:2018dnh}. 
The 2-3 mixing angle $\theta_{23}$ for the atmospheric neutrino oscillations 
are nearly maximal $\sim 45^{\rm o}$,
and it has been questioned whether the angle is really $45^{\rm o}$
or the angle deviates from it to
the higher or lower octant.
The current central value of $\theta_{23}$ for the global analysis~\cite{Esteban:2018azc,Capozzi:2019vbz}
is in the higher octant.
The measurement of the Dirac CP phase $\delta$ in the neutrino oscillations 
is important since it may tell us
something about the lepton number generation in the early universe.
The current measurements imply a large CP violation,
$\delta \sim - 90^{\rm o}$.
The accurate measurements of them will be one of the most important issues
in the next decade.

The 1-3 neutrino mixing angle $\theta_{13}$ has been measured accurately
by reactor neutrino oscillations \cite{An:2015rpe},
and the angle $\theta_{13} \simeq 8^{\rm o}-9^{\rm o}$ is much smaller than the other two,
 $\theta_{12} \sim 34^{\rm o}$, $\theta_{23} \sim 45^{\rm o}$.
When the (nearly) maximal atmospheric neutrino mixing is revealed at Super-Kamiokande,
the 1-3 neutrino mixing has been bounded from above by reactor 
neutrino at CHOOZ \cite{Apollonio:1999ae}.
The $\mu$-$\tau$ exchange symmetry has been considered
to realize such pattern of the neutrino mixings~\cite{Fukuyama:2017qxb,Xing:2015fdg}.
Under the $\mu$-$\tau$ exchange symmetry, i.e., if the neutrino mass matrix has a symmetry
under the $\nu_\mu$-$\nu_\tau$ exchange,
the 2-3 mixing is maximal and the 1-3 mixing is zero (the 1-2 mixing is free).
Surely, the observed 1-3 mixing is not zero,
and the $\mu$-$\tau$ symmetry is an approximate symmetry.
We claim that the separation of $\mu$-$\tau$ symmetric and $\mu$-$\tau$ breaking pieces
is a good parametrization
of the neutrino mass matrix
to describe the deviation from the maximal angle of the 2-3 mixing ($\delta \theta_{23}$), 1-3 mixing $\theta_{13}$, and 
the CP phase $\delta$
(even if there is not an underlying $\mu$-$\tau$ symmetry in the Lagrangian).

If there is $\mu$-$\tau$ exchange symmetry 
and the symmetry is spontaneously broken in the neutrino sector,
the deviation $\delta \theta_{23}$ is the similar size of $\theta_{13}$
very naively.
In this sense, the observed 1-3 mixing angle is too large
to explain the nearly maximal angle $\theta_{23}$,
which may be the reason why people have fewer interests on the $\mu$-$\tau$ symmetry now.
However, 
the small deviation $\delta \theta_{23}$ with the relatively large size of $\theta_{13}$
suggests that the CP violation in the neutrino sector is large.
Due to this consciousness, 
we will work on the description of the $\mu$-$\tau$ exchange symmetry and its breaking
in this paper.

We first describe
the neutrino mixings using the parametrization to separate the
$\mu$-$\tau$ symmetric and breaking pieces of the neutrino mass matrix.
If the $\mu$-$\tau$ breaking parameters are free in general,
any values of $\delta \theta_{23}$ and the CP phase $\delta$ is possible
obviously due to the number of parameters.
However, one can recognize that a large phase in the lepton sector is 
preferable to suppress $\delta \theta_{23}$.
The separation of the $\mu$-$\tau$ symmetric and breaking pieces 
is useful to understand this feature.
We exhibit several relations among 
$\delta \theta_{23}$, $\theta_{13}$ and the CP phase $\delta$
by special conditions for the $\mu$-$\tau$ breaking parameters.
It will be important to test the relations 
in the future when $\delta \theta_{23}$ and the CP phase $\delta$ are
accurately measured,
and to decode the underlying physics which determines the neutrino oscillation parameters.

We next construct a spontaneous $\mu$-$\tau$ symmetry breaking model,
which can restrict the $\mu$-$\tau$ breaking parameters.
In the model,
the $\mu$-$\tau$ exchange symmetry is broken
in a hidden sector,
and extra gauge interactions mediate the symmetry breaking to the neutrino sector.
Though one can easily build models in which the scalar fields to break 
the symmetry can couple with neutrino fields directly,
the $\mu$-$\tau$ breaking parameters in such models can be anything
by an artificial choice of the Yukawa-type interaction.
In our model, on the other hands, 
the $\mu$-$\tau$ breaking parameters are related to the extra gauge boson masses
and the standard model (SM) singlet neutrino mass spectrum,
and thus, the physical meaning of the $\mu$-$\tau$ breaking parameters is clearer.
We introduce so-called $B-L$ gauge boson for $\mu$-$\tau$ even gauge interaction,
and $L_\mu - L_\tau$ gauge boson for $\mu$-$\tau$ odd gauge interaction.
The mixings of those two gauge bosons are generated by the spontaneous breaking
of the $\mu$-$\tau$ symmetry,
and induce the $\mu$-$\tau$ breaking in the neutrino mass matrix by the gauge boson loops.

\medskip

This paper is organized as follows:
In Section 2, we describe the separation of the $\mu$-$\tau$ symmetric and breaking pieces
in the neutrino mass matrix,
and we discuss how the large CP phase is preferred exhibiting the relations among $\delta\theta_{23}$,
$\theta_{13}$ and $\delta$.
In Section 3, we build a model of the spontaneous $\mu$-$\tau$ exchange symmetry breaking.
Section 4 is devoted to the conclusion of this paper.
In Appendix A, we show the procedures of the diagonalization of the neutrino mass matrix
in the basis where $\mu$-$\tau$ symmetric and breaking pieces are separated,
in which the relations among the neutrino oscillation parameters are derived.
In Appendix B, we discuss the relation of our description with the $\mu$-$\tau$ reflection symmetry.

\section{Neutrino mixing angles and $\mu$-$\tau$ breaking}

The $\mu$-$\tau$ symmetric 
neutrino mass matrix is given as\footnote{
By changing the sign of the third generation neutrino field,
the (1,3) elements can be changed to be $-A$ (and $C$ to be $-C$ as well).
The freedom of such field redefinition is surely unphysical. 
In this paper, 
the $\mu$-$\tau$ symmetric mass matrix is given as
 $(M_\nu)_{12}= (M_\nu)_{13}$, $(M_\nu)_{22} = (M_\nu)_{33}$
 as a convention.
}
\begin{equation}
M_\nu^0 = \left(
\begin{array}{ccc}
D & A & A \\
A & B & C \\
A & C & B
\end{array}
\right),
\label{Mnu0}
\end{equation}
in the basis that the charged-lepton mass matrix is diagonal.
We define a unitary matrix for 2-3 block rotation as
\begin{equation}
U^{(23)}(\theta,\phi) = \left(
 \begin{array}{ccc}
  1 & 0 & 0 \\
  0 &   \cos \theta & \sin\theta e^{-i\phi}  \\
  0 &  -\sin \theta e^{i \phi} & \cos\theta 
 \end{array}
\right),
\end{equation}
and $U^{(12)},U^{(13)}$ for 1-2 and 1-3 rotation unitary matrices similarly.
We obtain
\begin{equation}
\bar M_\nu^0 \equiv 
U_0^T M_\nu^0 U_0=
\left(
 \begin{array}{ccc}
  D & \sqrt2 A & 0 \\
  \sqrt2 A & E & 0 \\
  0 & 0 & F
 \end{array}
\right),
\qquad
U_0 = U^{(23)}\left(-\frac{\pi}4,0\right),
\label{Mnubar}
\end{equation}
where $E = B+C$ and $F= B-C$.
The matrix $\bar M_\nu^0$ can be diagonalized by $U^{(12)} (\theta_{12}, \phi)$
and the Pontecorvo-Maki-Nakagawa-Sakata (PMNS) neutrino mixing matrix
can be written as
\begin{equation}
U_{\rm PMNS} = U^{(23)} \left(- \frac{\pi}4,0\right) U^{(12)} (\theta_{12},\phi) .
\label{PMNS0}
\end{equation}
One finds that the 2-3 neutrino mixing is maximal and the 1-3 mixing is zero
under the $\mu$-$\tau$ symmetry.
Thus, the matrix is good to use a base to describe the observed neutrino mixings.
The $\mu$-$\tau$ breaking piece can be parametrized as
\begin{equation}
M_\nu^\prime = 
\left(
\begin{array}{ccc}
0 & -A' & A' \\
-A' & -B' & 0 \\
A' & 0 & B'
\end{array}
\right),
\label{Mnuprime}
\end{equation}
and 
the neutrino mass matrix $M_\nu = M_\nu^0 + M_\nu^\prime$.
The $\mu$-$\tau$ breaking piece can be written by the same 2-3 rotation as
\begin{equation}
\bar M_\nu^\prime = U_0^T M_\nu^\prime U_0 = 
\left(
\begin{array}{ccc}
0 & 0 & \sqrt2 A' \\
0 & 0 & B' \\
\sqrt2 A' & B' & 0
\end{array}
\right).
\label{Mnuprime}
\end{equation}
The deviation from $45^{\rm o}$ of the 2-3 mixing and
1-3 mixing are generated by $A'$ and $B'$.
Obviously, one can understand that the separation of the $\mu$-$\tau$ symmetric 
and breaking pieces is just parametrization of the neutrino mass matrix elements
and any values of mixings and Dirac CP phase
 are possible if one does not assume anything on $A'$ and $B'$.

Suppose that there is underlying $\mu$-$\tau$ symmetry
and the symmetry is broken,
and $A'$ and $B'$ are the same order very naively.
In this case, one expects\footnote{We surely use the Particle Data Group (PDG) 
convention \cite{Tanabashi:2018oca} to describe the mixing angles in the mixing matrix; 
the mixing angles are put in the first quadrant by unphysical field redefinition if they are not there.}
\begin{equation}
\delta \theta_{23}  \equiv \theta_{23} - 45^{\rm o} \approx \pm\theta_{13}.
\end{equation}
Namely, the non-zero 1-3 mixing angle implies that the 2-3 mixing angle is deviated from $45^{\rm o}$.
The observed 1-3 mixing angle is $8^{\rm o}-9^{\rm o}$
and $\theta_{23}$ is $41^{\rm o} - 51^{\rm o}$ for $3\sigma$ range under the current global fit \cite{Esteban:2018azc,Capozzi:2019vbz}.
The current global best fit for the 2-3 mixing by NuFIT4.1 \cite{Esteban:2018azc} is $\theta_{23} = (48.6^{+1.0}_{-1.4})^{\rm o}$.
We expect more precise measurements of the 2-3 mixing angle to see if $\delta \theta_{23}$ is non-zero in the up-coming experimental data
to distinguish neutrino models.

It is worth to describe how $\delta \theta_{23}$ and $\theta_{13}$
(and the Dirac CP phase $\delta$, as well) are generated from $A'$ and $B'$.
The description depends on the neutrino mass hierarchy, so-called normal hierarchy (NH) and inverted hierarchy (IH).
In NH,
$\theta_{13}$ is basically generated by $A'$ and $\delta \theta_{23}$ is generated by $B'$.
In IH, on the other hand, either $A'$ or $B'$ can generate both $\theta_{13}$ and $\delta \theta_{23}$.
The detail descriptions are given in Appendix A.
Since there are two complex parameters $A'$ and $B'$, there is no rigid relation among $\delta \theta_{23}$, $\theta_{13}$
and the Dirac CP phase $\delta$.
We here assume special conditions on $A'$ and $B'$ to express a relation among them.
Derivation of the relations are given in Appendix A.

\medskip

$\bullet$ NH

We demand a condition that $A' \sim B'$.
We suppose that the
(1,3) element of the $\mu$-$\tau$ breaking matrix is zero
after $M_\nu^0$ is diagonalized by Eq.(\ref{PMNS0}).
This condition is satisfied if $\mu$-$\tau$ symmetric matrix $M_\nu^0$ is rank 2,
and the neutrino mass matrix is rank 2 even after the $\mu$-$\tau$ breaking term is added.
Then, we obtain\footnote{
 We note that Eq.(\ref{eq8}) holds if both $M_\nu^0$ and $M_\nu^0 + M_\nu^\prime$
 are rank 2, model-independently.
 Therefore, for example, it can hold 
 in the minimal seesaw model (e.g., numerical calculations are
 found in 
 \cite{Shimizu:2017fgu})
with $\mu$-$\tau$ reflection symmetry breaking by renormalization group evolution
\cite{Nath:2018hjx}. 
 A similar relation can hold in a neutrino model with $Z_2$ symmetry to realize $A'\sim B'$
 \cite{Ge:2011ih}.
}
\begin{equation}
 \theta_{23} - \frac{\pi}4 \simeq  \cot \theta_{12} \sin\theta_{13} \cos\delta.
 \label{eq8}
\end{equation}

$\bullet$ IH

In this case, either $A'$ or $B'$ can generate both $\theta_{13}$ and $\delta \theta_{23}$.
In order to have a rigid relation, we just assume that one of $A'$ and $B'$ is zero 
to avoid their contributions to be cancelled simply.
We obtain
\begin{eqnarray}
A' = 0 \qquad &\rightarrow& \qquad \theta_{23} - \frac{\pi}4  \simeq \cot 2\theta_{12} \sin\theta_{13} \cos\delta,
\label{eq9} \\
B' = 0 \qquad &\rightarrow& \qquad \theta_{23} - \frac{\pi}4 \simeq -\tan 2\theta_{12} \sin\theta_{13} \cos\delta.
\label{eq10}
\end{eqnarray}

\medskip

It is interesting to remark that the Dirac CP phase $\delta$ needs to be large ($|\cos\delta|$ needs to be small)
to satisfy the range of $\theta_{23}$ for the current global fit in Eqs.(\ref{eq8}) and (\ref{eq10})
(for $|\delta \theta_{23}| \alt 6^{\rm o}$, one obtains $|\cos\delta| \alt 0.45,0.3$ for Eqs.(\ref{eq8}),(\ref{eq10}), respectively).
Small change for a large phase can be
  simply analogized to $|1+z|^2 = 1 + 2\, {\rm Re}\, z + |z|^2$, namely,
$|1+z|^2 = 1 + O(|z|^2)$ for $|z| \ll 1$ if arg$(z)$ is nearly $\pm \pi/2$,
while $|1+z|^2 = 1\pm2 |z|+ O(|z|^2)$ for arg$(z) \simeq 0$ or $\pi$.
The description in Appendix A to derive the relations 
makes the meaning of this analogy clearer.
As a result, if the neutrino mass matrix has $\mu$-$\tau$ symmetry
and the symmetry is violated by $A' \sim B'$,
it is preferred to have a large phase to keep $|\delta \theta_{23}| < \theta_{13}/2$ roughly.
Of course, in general, the large phase is not necessarily same as the Dirac CP phase
which can be measured by the neutrino oscillations and/or $\delta \theta_{23}$ can be cancelled
irrespective of the phase.
If special cases are considered as above,
the phase really corresponds to the Dirac CP phase, and those simple relations are satisfied.

It is also interesting to note that
one needs $\cos\delta > 0$ $(<0)$ if $\theta_{23}$ is in the higher (lower) octant 
if the relations are given as in Eqs.(\ref{eq8}) and (\ref{eq9}).
If $\delta \theta_{23} = 3^{\rm o}$ as given by the current central value,
one obtains 
$|\delta| = 76^{\rm o}$, $32^{\rm o}$, and 
$98^{\rm o}$ for Eqs.(\ref{eq8}), (\ref{eq9}), and (\ref{eq10}), respectively.
It is interesting to
test the relations and decode the $A',B'$ parameters 
if $\delta \theta_{23}$ and $\delta$ are measured more accurately.

We emphasize that the CP phase is naively preferred to be large
if the neutrino mass matrix has approximate $\mu$-$\tau$ symmetry,
and $\theta_{13}$ is generated by spontaneous $\mu$-$\tau$ breaking,
though the relation among $\theta_{13}$, $\delta \theta_{23}$ and $\delta$
cannot be rigid in general.
It is worth to construct a $\mu$-$\tau$ symmetry breaking model.

\section{A model for $\mu$-$\tau$ symmetry breaking}

We have studied that
the description of $\mu$-$\tau$ symmetric and $\mu$-$\tau$ breaking pieces
is a good base
to consider the size of $\delta \theta_{23}$ and the Dirac CP phase $\delta$,
which will be measured more accurately at the near future long-baseline
neutrino oscillation experiments.
In this section, we will construct a model for the $\mu$-$\tau$ breaking.

\subsection{Spontaneous breaking of $\mu$-$\tau$ exchange symmetry}

In the beginning, let us consider a set of two complex scalars, named as $(\phi_\mu, \phi_\tau)$,
and build a scalar potential, which has $\phi_\mu \leftrightarrow \phi_\tau$ exchange symmetry.
Suppose that $\phi_\mu$ and $\phi_\tau$ have charges of a U(1) symmetry.
Then, the potential can be written as
\begin{equation}
V = - m^2 (|\phi_\mu|^2 +|\phi_\tau|^2) + \lambda_1 (|\phi_\mu|^4 + |\phi_\tau|^4) + \lambda_2 |\phi_\mu|^2 |\phi_\tau|^2.
\end{equation}
One finds the minimum of the potential,
\begin{equation}
\phi_\mu^2 = \phi_\tau^2 = \frac{m^2}{2\lambda_1+\lambda_2},
\end{equation}
for $2\lambda_1 + \lambda_2 >0$,
and 
\begin{equation}
\phi_\mu^2 =\frac{m^2}{2\lambda_1},\  \phi_\tau = 0, \qquad {\rm or}
\qquad
\phi_\mu = 0, \ \phi_\tau^2 =  \frac{m^2}{2\lambda_1}, 
\end{equation}
for $2\lambda_1 + \lambda_2 < 0$.
The former keeps the exchange symmetry,
and the latter violates it spontaneously.
If the $\mu$-$\tau$ symmetric vacuum expectation value (vev) and $\mu$-$\tau$ breaking vev
directly couples to neutrino sector,
one can easily construct a model with discrete symmetry such as $S_3$ \cite{Hall:1995es,Harrison:2003aw,Chen:2004rr,Dev:2012sg,Pramanick:2019oxb}.

We will construct a model in which the $\mu$-$\tau$
exchange symmetry is broken in a hidden sector,
namely the scalars ($\phi_\mu,\phi_\tau$)
does not couple to neutrinos directly,
and the violation is mediated by a gauge interaction.
We introduce gauge bosons which are even and odd under the exchange symmetry,
and the mixing of the gauge bosons can generate the $\mu$-$\tau$ breaking
in the neutrino mass matrix.
Though the easy candidate of the $\mu$-$\tau$ even gauge boson is $Z$-boson,
the sizable mixing of the $Z$ boson with the other gauge boson to generate $\theta_{13}$
contradicts with the precision measurements.
Therefore, we consider $B-L$ gauge boson for the $\mu$-$\tau$ even gauge boson.
More precisely speaking, the hypercharge is a linear combination of $U(1)_{B-L}$ and $U(1)_R$,
and the extra $\mu$-$\tau$ even gauge boson we call as $Z'$ is the one for $U(1)'$ symmetry
which is 
orthogonal to the hypercharge.
The $\mu$-$\tau$ odd gauge boson, we call it as $Z''$, is a U(1) gauge boson
for $L_\mu - L_\tau$ charge.
Namely, the gauge interactions to the lepton doublets are written as
\begin{eqnarray}
{\cal L} &\supset& i g' Z' (\bar\ell_e \gamma \ell_e + \bar\ell_\mu \gamma \ell_\mu + \bar\ell_\tau \gamma \ell_\tau) \qquad (\mu-\tau\  {\rm even}) \\
{\cal L} &\supset& i g'' Z'' ( \bar\ell_\mu \gamma \ell_\mu - \bar\ell_\tau \gamma \ell_\tau)
\qquad \qquad \quad (\mu-\tau\  {\rm odd})
\end{eqnarray}
where $g'$ and $g''$ are gauge couplings ($g'$ is not a hypercharge gauge coupling $g_Y$ for our notation).

Suppose that
the three scalars $\phi_\mu$, $\phi_\tau$ and $\phi$ has
$U(1)'$ and $L_\mu - L_\tau$ charges as given in Table~1.
\begin{table}[t]
\begin{center}
\begin{tabular}{|c|ccc|} \hline
& $\phi_\mu$ & $\phi_\tau$ & $\phi$ \\ \hline
$U(1)'$ & 1 & 1 & 1 \\
$L_\mu-L_\tau$ & $1$ & $-1$ & 0 \\ \hline
\end{tabular} 
\end{center}
\caption{The U(1) charges of the scalar fields.}
\end{table}
The gauge boson mass term is obtained as
\begin{equation}
\left(\begin{array}{cc}
Z' & Z''
\end{array}\right)
\left(\begin{array}{cc}
g'^2 (\phi^2 + \phi_\mu^2 + \phi_\tau^2) & g' g'' (\phi_\mu^2 - \phi_\tau^2) \\
g' g'' (\phi_\mu^2 - \phi_\tau^2) & g''^2 (\phi_\mu^2 + \phi_\tau^2)
\end{array}\right)
\left(\begin{array}{c}
Z' \\ Z''
\end{array}\right).
\end{equation}
Obviously, the $Z'$-$Z''$ mixing is absent in the $\mu$-$\tau$ symmetric vacua
($\phi_\mu = \phi_\tau$)
and they are mixed in the $\mu$-$\tau$ breaking vacua.
Suppose $\phi_\mu = 0$ and $\phi_\tau \neq 0$ for the $\mu$-$\tau$ breaking vacua,
then we obtain the $Z'$-$Z''$ mixing angle $\alpha$ as
\begin{equation}
\tan2\alpha = \frac{2g'g''\phi_\tau^2}{g'^2 \phi^2 + (g'^2-g''^2) \phi_\tau^2}.
\end{equation}
One finds that they are maximally mixed
if $g' = g''$ and $\phi\ll \phi_\tau$.
Surely, there can be additional scalar fields to break the U(1) symmetries.
The request of the model will be that 
 the $Z'$-$Z''$ mixing is not too small to generate the 1-3 neutrino mixing from the $\mu$-$\tau$ breaking.

We comment that gauge kinetic mixing between $Z'$ and $Z''$ is possible in general.
We can suppose that the kinetic mixing is absent by the $\mu$-$\tau$ exchange symmetry
(since $Z'$ is even and $Z''$ is odd) and the $Z'$-$Z''$ mixing is generated by the
spontaneous breaking of $\mu$-$\tau$ symmetry.

\subsection{Loop-induced neutrino mass}

We build a model where the $\mu$-$\tau$ breaking in the neutrino mass matrix is generated
via the gauge boson loop.
If there is a tree-level neutrino mass, however,
the loop correction is tiny 
and the observed size of $\theta_{13}$ cannot be generated unless the gauge interaction is so strong.
Therefore, 
we consider a situation that the $\mu$-$\tau$ symmetric neutrino mass
is also generated by a loop effect.
In order to realize such setup, we consider the following neutrino mass term (for one generation),
\begin{equation}
\left( \begin{array}{ccc}
\nu & N & S 
\end{array}
\right)
\left( \begin{array}{ccc}
0 & 0 & m \\
0 & 0 & X \\
m & X & M 
\end{array}
\right)
\left( \begin{array}{c}
\nu \\ N\\ S 
\end{array}
\right),
\end{equation}
and study how the loop-induced neutrino mass is generated.
We remark that a neutrino, which is mainly active neutrino $\nu$, is massless at the tree-level as long as $N$-$N$ and $\nu$-$N$ elements
are zero.
Surely, those elements are not necessarily exactly zero.
The tree-level neutrino masses are supposed to be less than meV (if there are) as a setup.
The active neutrino mass, to explain the observed neutrino oscillations, can be generated radiatively even if the tree-level mass is zero.
Since we will gauge the $B-L$ symmetry, we assign $\nu \subset \ell$ has $B-L= -1$,
and $0$, $+1$ for $N$, $S$, respectively.

%

The induced neutrino mass via $Z$-boson and Higgs boson loops
is \cite{Pilaftsis:1991ug}
\begin{equation}
m_\nu = \frac{\alpha_2}{16\pi} \sum_{k = 1,2}
\frac{M_k}{M_W^2} C_{\nu N_k}^2 
\left[
M_k^2 (f(M_k,M_Z) - f(M_k, M_H) ) - 4 M_Z^2 f(M_k,M_Z)
\right],
\label{loop-induced}
\end{equation}
where
$M_k$ is the mass of the mass eigenstate $N_k$
and $C_{\nu N_k}$ is the mixing (which is $Z$-$\nu$-$N_k$ coupling).
The loop function $f$
is given as
\begin{equation}
f(M_N,M) = \frac{M_N^2}{M_N^2 - M^2} \ln \frac{M_N^2}{M^2}
+ \ln \frac{M^2}{Q^2} -1.
\end{equation}
The loop-induced mass does not depend on the renormalization scale $Q$ after all.
The first term in the bracket in Eq.(\ref{loop-induced}) corresponds to the contribution from the loop diagrams for 
Higgs and Nambu-Goldstone modes.
These contributions are canceled if the singlet neutrino masses 
are much heavier than them, $M_i \gg M_{Z,H}$.
The second term corresponds to the $Z$ boson loop, which we will calculate hereafter.

The diagonalization matrix can be written as
\begin{equation}
U^T
= \left(
 \begin{array}{ccc}
  1 & 0 & 0 \\
  0 & c_R & - s_R \\
  0 & s_R & c_R
 \end{array}
\right)
\left(
 \begin{array}{ccc}
  c_\theta & -s_\theta & 0 \\
  s_\theta & c_\theta &0   \\
  0 & 0 & 1
 \end{array}
\right)
=\left(
 \begin{array}{ccc}
  c_\theta & -s_\theta & 0 \\
  c_R s_\theta & c_R c_\theta & - s_R  \\
  s_R s_\theta & s_R c_\theta & c_R
 \end{array}
\right),
\end{equation}
with an obvious notation, $s_R = \sin\theta_R$, $s_\theta = \sin\theta$, etc,
and
\begin{equation}
\left( \begin{array}{c}
 \nu \\ N \\ S
\end{array}
\right)
=
U \left( \begin{array}{c}
 \nu_1 \\ N_1 \\ N_2
\end{array}
\right),
\end{equation}
where $\nu_1$, $N_i$ are mass eigenstates.
We obtain
\begin{equation}
\tan\theta = \frac{m}{X},
\qquad
\tan2\theta_R = \frac{2\sqrt{X^2+m^2}}{M}.
\end{equation}
 We have a relation
 \begin{equation}
c_R^2 M_1 + s_R^2 M_2 = 0,
\end{equation}
which corresponds to the condition that the tree-level seesaw neutrino mass is zero.
The $Z-\nu-N_k$ coupling are given as
\begin{equation}
C_{\nu N_1} = c_\theta s_\theta c_R,
\qquad
C_{\nu N_2} = c_\theta s_\theta s_R.
\end{equation}
%
The neutrino mass from $Z$ boson loop is then
\begin{eqnarray}
m_\nu &=& - \frac{\alpha_Z}{4\pi} (s_\theta c_\theta)^2 (M_1 c_R^2 f(M_1,M_Z) +  M_2 s_R^2 f(M_2,M_Z) )
\\
&=& - \frac{\alpha_Z}{4\pi} (s_\theta c_\theta)^2 M_1 c_R^2 (f(M_1,M_Z) - f(M_2, M_Z)) \\
&=& -\frac{\alpha_Z}{4\pi} (s_\theta c_\theta)^2 M_1 c_R^2 \left(
 \frac{M_1^2}{M_1^2 - M_Z^2} \ln \frac{M_1^2}{M_Z^2}
 - \frac{M_2^2}{M_2^2 - M_Z^2} \ln \frac{M_2^2}{M_Z^2}
\right),
\end{eqnarray}
where $\alpha_Z = (g_2^2+g_Y^2)/(4\pi)$.
We note
that the loop-induced neutrino mass tends to be zero for a limit $M/X \to 0$
($M_1 \simeq -M_2$).
This can be understood as that a (global) lepton number symmetry remains for $M = 0$.

We emphasize that the loop-induced neutrino mass does not depend on $Z$ boson mass,
if the singlet neutrinos are heavy.
In fact, one finds
\begin{equation}
m_\nu = -\frac{\alpha_Z}{4\pi} (s_\theta c_\theta)^2 M_1 c_R^2 \ln \frac{M_1^2}{M_2^2},
\end{equation}
for $M_1,M_2 \gg M_Z$.
Therefore, the contributions from $Z'$ and $Z''$ gauge bosons
are the same order as the one from $Z$ boson if they are lighter than the heavy neutrino mass and gauge couplings are the same size of the electroweak coupling.
The neutrino mass $M$ violates the $B-L$ symmetry and the $B-L$ gauge boson and 
the neutrino mass may be similar size.
Using the freedom of the mass spectrum of gauge bosons and singlet neutrinos,
we will control the induced $\mu$-$\tau$ symmetric and 
violating terms.

Now we understand the loop-induced neutrino mass for one-generation, 
and it can be extended to the three-generation version.
The neutrino mass terms are given as
\begin{equation}
\left( \begin{array}{ccc}
\nu & N & S 
\end{array}
\right)
\left( \begin{array}{ccc}
0 & 0 & m \\
0 & 0 & X \\
m^T & X^T & M 
\end{array}
\right)
\left( \begin{array}{c}
\nu \\ N\\ S 
\end{array}
\right),
\end{equation}
where each element is $3\times 3$ matrix.
We assume that the active and singlet neutrino mixings are small ($m X^{-1}$ is small).
Then,
by extending the same process to obtain the loop-induced neutrino mass for the 1-generation case,
 the neutrino mass matrix for 3-generation case from $Z$-boson loop is approximately given as
\begin{equation}
M_\nu = -\frac{\alpha_Z}{4\pi} 
\sum_{a = 1,2} m X^{-1} U_a [M^{(a)} f(M^{(a)},M_Z)] U_a^T X^{-1}{}^T m^T.
\label{Zloop}
\end{equation}
Here, $ [M^{(a)} f(M^{(a)},M_Z)] $ denotes a diagonal matrix:
$[A] \equiv {\rm diag}.(A_1,A_2,A_3)$,
%
and
\begin{equation}
\left(\begin{array}{cc}
 U_1^\dagger & U_3^\dagger  \\
 U_2^\dagger & U_4^\dagger
\end{array}
\right)
\left(\begin{array}{cc}
 0 & X  \\
 X^T & M
\end{array}
\right)
\left(\begin{array}{cc}
 U_1^* & U_2^*  \\
 U_3^* & U_4^*
\end{array}
\right)
= 
\left(\begin{array}{cc}
 M^{(1)} & 0  \\
 0 & M^{(2)}
\end{array}
\right),
\label{U1U2}
\end{equation}
and $M^{(1,2)}$ are $3\times 3$ diagonal matrices.
%
By definition,
\begin{equation}
U_1 M^{(1)} U_1^T + U_2 M^{(2)} U_2^T = 0.
\end{equation}
In the following, we denote the $Z$-loop-induced neutrino mass factoring out the gauge coupling
as
\begin{equation}
M_\nu = g_Z^2 \hat M(M_Z).
\end{equation}
The loop-induced neutrino mass depends on the heavy singlet neutrino masses explicitly,
but we omit the dependence to write the equations shortly.
%
We assume that tree-level term has $\mu$-$\tau$ exchange symmetry,
and then, the matrix $\hat M_\nu (M_Z)$ has the symmetry.

\subsection{Loop-induced $\mu$-$\tau$ breaking in the neutrino mass matrix}

The gauge interaction of the extra gauge bosons to the leptons can be written
as
\begin{equation}
i \bar\ell_i (g^\prime Z^\prime_\mu I_{ij} + g'' Z''_\mu T_{ij} ) \gamma^\mu \ell_j,
\end{equation}
where $I$ is an identity matrix and
\begin{equation}
T = 
\left(\begin{array}{ccc}
 0 & &   \\
 & 1 & \\
 &&-1
\end{array}
\right).
\label{T}
\end{equation}
%
The gauge bosons are mixed, and the mass eigenstates $Z_1$ and $Z_2$ are written as
\begin{equation}
Z_1 = c_\alpha Z' + s_\alpha Z'',\qquad
Z_2 = -s_\alpha Z' + c_\alpha Z''.
\end{equation}
Then, the gauge interaction is written as
\begin{equation}
i \bar\ell_i ((c_\alpha g^\prime  I_{ij} + s_\alpha g'' T_{ij}) Z_1^\mu + (-s_\alpha g^\prime  I_{ij} + c_\alpha g'' T_{ij}) Z_2^\mu  ) \gamma_\mu \ell_j.
\end{equation}
We obtain 
the loop-induced neutrino mass
as
\begin{eqnarray}
M_\nu &=& 
g_Z^2 \hat M(M_Z) \nonumber
+ (c_\alpha g^\prime  I + s_\alpha g'' T)  \hat M(M_{Z_1}) (c_\alpha g^\prime  I + s_\alpha g'' T)  \\
&&+ (-s_\alpha g^\prime  I + c_\alpha g'' T) \hat M (M_{Z_2}) (-s_\alpha g^\prime  I + c_\alpha g'' T) 
\\
&=&
g_Z^2 \hat M(M_Z)\nonumber
+
g'^2 c_\alpha^2 \hat M(M_{Z_1}) + g'^2 s_\alpha^2 \hat M(M_{Z_2}) \\
&&+ \nonumber
g''^2 s_\alpha^2 T \hat M(M_{Z_1}) T + g''^2 c_\alpha^2 T \hat M(M_{Z_2})T \\
&&+ c_\alpha s_\alpha g' g'' \left( T ( \hat M(M_{Z_1}) -\hat M(M_{Z_2})) +
 ( \hat M(M_{Z_1}) -\hat M(M_{Z_2}) )T \right).
 \label{loop-induced-mass}
\end{eqnarray}
The last term is the $\mu$-$\tau$ breaking term.
As we have remarked,
$\hat M$ does not depend on the gauge boson mass
if the singlet neutrinos are much heavier than the gauge boson.
Therefore, in that case, the $\mu$-$\tau$ breaking term vanishes.


In order to calculate the loop-induced neutrino mass matrix expressed 
in Eq.(\ref{loop-induced-mass})),
one needs to diagonalize the $6\times 6$ heavy neutrino mass matrix as given in Eq.(\ref{U1U2}).
The diagonalization can be surely done numerically in any setup.
We here assume a setup to express the loop-induced mass matrix in a simple form
and to illustrate the essence of the structure.
We assume that the singlet neutrino mass matrices $X$ and $M$ are also $\mu$-$\tau$ symmetric.
We note that the $L_\mu - L_\tau$ symmetry breaking vev can be directly applied to the SM singlet masses,
but the $L_\mu - L_\tau$ breaking in the Dirac mass matrix $m$ will be higher order\footnote{
We comment that the $L_\mu-L_\tau$ charges for the lepton doublets are assigned as
$0, 1, -1$, and those of the right-handed leptons are $0, -1, 1$. The charged lepton mass matrix is then diagonal up to the higher order terms.
}.
Due to the $\mu$-$\tau$ symmetry,
one can suppose that
the matrices are given after 2-3 generation are rotated by $\pi/4$ as
\begin{equation}
\bar m=\left(\begin{array}{ccc}
 d& a& 0 \\
 a' & e & 0\\
 0& 0 & f
\end{array}
\right),\quad
\bar X=\left(\begin{array}{ccc}
 X_1& X_4& 0  \\
 X_5 & X_2 & 0 \\
 0&  0 &X_3 
\end{array}
\right),\quad
\bar M=\left(\begin{array}{ccc}
 M_1& 0 &  0 \\
 0 &  M_2 & 0\\
 0 &0 & M_3
\end{array}
\right).
\end{equation}
The matrix $\bar M$ has been diagonalized by rotating 1-2 generation of the $S$ field,
and $a$, $a'$ in $\bar m$ are generated by the rotation. 
One can parameterize so that $X_4$ or $X_5$ are zero by rotating the $N$ field,
but both $\bar X$ and $\bar M$ cannot be diagonalized simultaneously in general.
In order to illustrate the induced mass matrix simply,
we here assume $X_4 = X_5 =0$
so that the diagonalization of the singlet neutrino mass matrix in Eq.(\ref{U1U2})
are split in each generation.
Then, we find that
the induced neutrino mass matrix 
can be written as
\begin{eqnarray}
\bar M_\nu &=&
{\cal M}_1 F_1 (X_1, M_1) +
{\cal M}_2 F_1 (X_2, M_2) + 
{\cal M}_3 F_1 (X_3, M_3) \\ 
&&+ 
{\cal M}_4 F_2 (X_1, M_1) +
{\cal M}_5 F_2 (X_2, M_2) +
{\cal M}_6 F_2 (X_3, M_3)
\nonumber \\
&&
+
{\cal M}_6 F_3 (X_1, M_1) +
{\cal M}_7 F_3 (X_2, M_2) +
{\cal M}_9 F_3 (X_3, M_3), 
\nonumber
\end{eqnarray}
where the matrices ${\cal M}_i$ are\footnote{
${\cal M}_1 = \bar m \, {\rm diag}(1,0,0)\,   \bar m^T$,
${\cal M}_2 = \bar m \, {\rm diag}(0,1,0)\,   \bar m^T$,
${\cal M}_3 = \bar m \, {\rm diag}(0,0,1)\,   \bar m^T$,
${\cal M}_4 = \bar T {\cal M}_1 \bar T$,
${\cal M}_5 = \bar T {\cal M}_2 \bar T$,
${\cal M}_6 = \bar T {\cal M}_3 \bar T$,
${\cal M}_7 = \bar T {\cal M}_1+ {\cal M}_1 \bar T$,
${\cal M}_8 = \bar T {\cal M}_2+ {\cal M}_2 \bar T$,
${\cal M}_9 = \bar T {\cal M}_3+ {\cal M}_3 \bar T$,
where $\bar T$ is the matrix for $\mu$-$\tau$ odd gauge interaction, Eq.(\ref{T}), in the basis
that the 2-3 generation is rotated by $\pi/4$,
\begin{equation}
\bar T  =\left(  
\begin{array}{ccc}
 0 & 0 & 0 \\
 0 & 0 & 1 \\
 0 & 1 & 0 
\end{array}
\right).
\end{equation}
}
\begin{eqnarray}
&\!\!&\!\!
{\cal M}_1=\left(\begin{array}{ccc}
 d^2 & a'd &   0 \\
 a'd &  a'^2 & 0 \\
 0 & 0 & 0
\end{array}
\right),
\quad
{\cal M}_2=\left(\begin{array}{ccc}
 a^2 & ae &   0 \\
 ae &  e^2 & 0 \\
 0 & 0 & 0
\end{array}
\right),
\quad
{\cal M}_3=\left(\begin{array}{ccc}
 0 & 0 &   0 \\
 0 &  0 & 0 \\
 0 & 0 & f^2
\end{array}
\right), \\
&\!\!&\!\!
{\cal M}_4=\left(\begin{array}{ccc}
 0 & 0 &   0 \\
 0 &  0 & 0 \\
 0 & 0 & a'^2
\end{array}
\right),
\qquad
{\cal M}_5=\left(\begin{array}{ccc}
 0 & 0 &   0 \\
 0 &  0 & 0 \\
 0 & 0 & e^2
\end{array}
\right),
\qquad
{\cal M}_6=\left(\begin{array}{ccc}
 0 & 0 &   0 \\
 0 &  f^2 & 0 \\
 0 & 0 & 0
\end{array}
\right), \\
&\!\!&\!\!
{\cal M}_7=\left(\begin{array}{ccc}
 0 & 0 &   a'd \\
 0 &  0 & a'^2 \\
 a'd & a'^2 & 0
\end{array}
\right),
\quad
{\cal M}_8=\left(\begin{array}{ccc}
 0 & 0 &   ae \\
 0 &  0 & e^2 \\
 ae & e^2 & 0
\end{array}
\right),
\quad
{\cal M}_9=\left(\begin{array}{ccc}
 0 & 0 &   0 \\
 0 &  0 & f^2 \\
 0 & f^2 & 0
\end{array}
\right).
\end{eqnarray}
Defining
\begin{equation}
F(X,M,M_z) =
 \frac{c_{R}^2}{X^2} M^{(1)} (f(M^{(1)} ,M_z) - f(M^{(2)} ,M_z)) ,
\end{equation}
where $c_R = \cos\theta_R$, $\tan2\theta_{R} = 2X/M$ and $M^{(1,2)} = (M \pm \sqrt{M^2 + 4X^2})/2$,
we give the functions as
\begin{eqnarray}
F_1 (X,M) &=& \frac1{16\pi^2} [ g_Z^2 F(X,M,M_Z) + g'^2 (c_\alpha^2 F(X,M,M_{Z_1}) + s_\alpha^2 F(X,M,M_{Z_2}) ) ], \\
F_2 (X,M) &=& \frac{1}{16\pi^2}  g''^2 (s_\alpha^2 F(X,M,M_{Z_1}) + c_\alpha^2 F(X,M,M_{Z_2})), \\
F_3 (X,M) &=& \frac{1}{16\pi^2}  g' g'' s_\alpha c_\alpha ( F(X,M,M_{Z_1}) - F(X,M,M_{Z_2})).
\end{eqnarray}
One finds that the matrices ${\cal M}_{7,8,9}$ correspond to the $\mu$-$\tau$ breaking
terms.
We list the property of the function $F(X,M,M_z)$, which is useful to control the neutrino mass hierachy:
\begin{enumerate}
\item
For $X,M \gg M_z$, the function $F$ is less dependent on $M_z$.
\item
For $X,M \ll M_z$, the function $F$ is suppressed by $\sim X^2/M_z^2, M^2/M_z^2$.
\item
For $M \ll X$, the function $F$ is suppressed as a result of a global lepton number symmetry.
\end{enumerate}

Now let us consider how the neutrino mixings are reproduced from ${\cal M}_i$.
Remind that they are written in the basis that the 2-3 generation is rotated by $\pi/4$.
The neutrino mass matrices are given as Eqs.(\ref{Mnubar}) and (\ref{Mnuprime}).
In the NH case,
the (3,3) element is the largest.
The simplest realization is that $f$ is the largest in ${\cal M}_3$.
The other elements are smaller than the (3,3) element
by the original Dirac Yukawa coupling for $m$ 
or by the spectrum of the singlet neutrinos.
For example, if $f > a,a',d,e$ in the matrix $\bar m$,
the choice of $X_3,M_3 < M_{Z_1}, M_{Z_2}$
can suppress $F(X_3,M_3,M_{Z_{1,2}})$
not to disturb the hierarchical structure by the ${\cal M}_{6,9}$ terms.
Surely, for $X_3, M_3 > M_Z$, ${\cal M}_3$ term is not suppressed.
Because the neutrino mass ratio $m_2/m_3 \approx \sqrt{\Delta m^2_{\rm sol}/\Delta m^2_{\rm atm}}$ 
is the same size as the 1-3 mixing
in NH,
$g_Z^2 \sim g' g'' s_\alpha c_\alpha$ is needed naively.
For another choice, we can also reproduce the mass hierarchy by the size of the gauge couplings:
the contribution from ${\cal M}_4$ or ${\cal M}_5$ is the largest.
This can be done if $L_\mu - L_\tau$ gauge coupling $g''$ is a bit ($2-3$ times) stronger than
$g_Z$ and $g'$.

Naively speaking, the large solar mixing is generated by $a'd$ and $ae$
in ${\cal M}_1$ and ${\cal M}_2$.
Then, if the 1-3 mixing is generated by the $\mu$-$\tau$ breaking terms, ${\cal M}_7$ and ${\cal M}_8$,
one can understand that the 2-3 mixing is also deviated from the maximal mixing due to $A' \sim B'$.
Therefore, as explained, a large CP phase in the components is preferable 
to keep the 2-3 mixing nearly maximal, 
though there can be freedom for cancellation between (2,3) components in
${\cal M}_{7,8}$ in general.
If the neutrino mass matrix is rank 2, such cancellation can be avoided.
The rank-2 mass matrix can be easily constructed:
for example, 
the ${\cal M}_{1,4,7}$ contributions are tiny if $M_1 \ll X_1$.
As we have remarked, $F(X_1,M_1,M)$ is small if $M_1 \ll X_1$.
Then, we have a relation given in Eq.(\ref{eq8}).
In this case, a large CP phase $\delta$ is definitely favored.
We comment that
the rank 2 condition is a sufficient one
to realize Eq.(\ref{eq8}).
One can find 
that
Eq.(\ref{eq8}) holds if $(d,a')$ and $(a,e)$ are ``orthogonal" (i.e. $d a^* + a' e^* = 0$),
supposing that the neutrino mass matrix is given as a linear combination
of ${\cal M}_{1,2,4,7}$.

In the case of IH,
one can build the neutrino mass matrix
so that the ${\cal M}_{1,2}$ contributions are dominant,
and the others are subdominant,
namely,
the $Z'$ and $Z''$ loop contributions are  $\sim$ 10\% 
compared to the $Z$ boson contribution.
Such a situation can be easily achieved if the gauge coupling is small, and/or $Z'$-$Z''$ mixing is small,
and/or $Z'$,$Z''$ masses are a bit larger than the singlet neutrino masses.
If one chooses ${\cal M}_{7,8}$ to be suppressed by $M_{1,2}, X_{1,2} \ll M_{Z_{1,2}}$
and ${\cal M}_9$ to be a dominant contribution for $\mu$-$\tau$ breaking,
a situation for $A' =0$ can be obtained, and then, Eq.(\ref{eq9}) is satisfied approximately.

\medskip

We comment about the scales of the singlet neutrino masses 
and the extra gauge boson masses.
If the singlet neutrinos are much heavier than $Z$-boson,
the induced neutrino mass is roughly given
as
\begin{equation}
m_\nu \approx \frac{\alpha_Z}{4\pi} m^2 \frac{M}{X^2},
\end{equation}
and therefore,
the neutrino mass scale is similar to the usual seesaw (or inverse seesaw) up to the loop factor.
For the Dirac mass $m \sim 100$ GeV, one finds $M\sim X \sim 10^{13}$ GeV
to obtain $m_\nu \sim 0.05$ eV.
If $M\sim X \sim 1$ TeV, one finds $m \sim 1 $ MeV.\footnote{
%
%
The experimental constraints for the singlet sterile neutrino particles which can be detected 
by the lepton number violating processes at LHC 
is that those masses are more than several hundred GeV (depending on the active-sterile neutrino mixings)
\cite{Atre:2009rg,Aad:2015xaa,Khachatryan:2016olu}.
%
%
We note that the process via $Z'$ boson based on the left-right symmetric model is also analyzed in \cite{Aad:2015xaa}. 
%
Our model surely works even if those masses are ultra-heavy as usual seesaw models,
and thus one can always avoid such experimental constraints.
However, it is possible that the singlet neutrinos are detected at the LHC near future
if the singlet neutrinos lie just above the experimental constraints.
Actually, the active-sterile neutrino mixing, $\theta_{\rm as}$, is tiny in the naive seesaw structure ($\theta_{\rm as} \sim \sqrt{m_\nu/M}$),
and thus the mixing is enlarged to detect the process at LHC.
To do that, the neutrino mass matrix needs to have a similar structure as the model we have shown irrespective of
the existence of the $\mu$-$\tau$ symmetry.
The active-sterile mixing is $\theta_{\rm as} \sim m/X$, which is independent of the observed active neutrino masses,
is constrained from the other observables \cite{Atre:2009rg}.
One always needs to care about the loop corrections in such models for the lepton number violation which can be detected at LHC.}
If the SM singlet neutrinos are lighter than $Z$ boson
and $Z'$, $Z''$ bosons are heavier than $Z$ boson,
the $\mu$-$\tau$ breaking in the neutrino mass matrix will be small.
If $Z_1$ and/or $Z_2$ are around O(100) GeV and the neutrinos are as light as them,  
it may be possible to build the $\mu$-$\tau$ breaking neutrino mass matrix for light SM singlet neutrinos
and to make the extra gauge boson contribute to generate lepton flavor non-universality \cite{Bifani:2018zmi,Crivellin:2015era,Kohda:2018xbc},
as long as they are allowed by experiments
though we do not survey the possibility in this paper.
The very light (MeV-scale) extra gauge boson may also have phenomenological interests \cite{Araki:2015mya,Adachi:2019otg}.
As described, the extra U(1) symmetries are broken by the field given in Table 1.
The field $\phi_\tau \neq 0$ (with $\phi_\mu = 0$) breaks $\mu$-$\tau$ exchange symmetry
and $U(1)'$ and $L_\mu- L_\tau$ symmetry is broken to a linear combination.
The remained symmetry is broken by the vev of $\phi$, which has a $B-L$ charge.
The $\mu$-$\tau$ breaking can occur at a high scale, 
and the vev of $\phi$ can be much lower than it.
Then, the $Z'$-$Z''$ mixing angle is determined by the gauge couplings: $\tan\alpha \simeq g''/g'$,
and $M_{Z_1} \ll M_{Z_2}$.
The SM singlet neutrinos can lie around the mass of $Z_1$.
In this situation, the proper size of 
the $\mu$-$\tau$ breaking in the neutrino mass matrix can be easily generated.
The neutrino mass hierarchy and the size of $\theta_{13}$ can be adjusted
by the singlet neutrino mass spectrum and the choice of the gauge couplings $g'$ and $g''$,
as we have described.
The mass scale of the remained U(1) breaking, namely $\sim M_{Z_1}$, can be around TeV scale, and there may have an opportunity
to be observed experimentally\footnote{
We comment on the experimental constraints of the extra neutral gauge boson masses,
though the purpose of this paper is not to pursuit the corner of the parameter space.
As we have mentioned, the mixing between $Z$ boson and $Z'/Z''$ bosons are restricted 
(that is why we have introduced two extra gauge bosons).
If the mixing of them are absent, the experimental bounds are not so stringent.
The strongest constraints in that case often come from neutrino trident production processes \cite{Altmannshofer:2014pba},
which needs to be considered when the extra gauge bosons lie below the weak scale as in the references \cite{Kohda:2018xbc,Araki:2015mya}.
The muon has extra U(1) charges in our model, and thus the trident processes constrain the extra gauge boson masses.
The mass bounds depends on the gauge couplings, and the masses should be larger than several hundred GeV
if the gauge coupling is roughly same size as the electroweak gauge couplings.
If the gauge coupling is smaller, the masses can be much lighter.
Belle II collaboration reports a new constraint of the extra gauge bosons \cite{Adachi:2019otg},
and it is expected that the bounds of the light gauge bosons will be updated. 
}.

Finally, we comment on the evolution by the renormalization group equations (RGE).
In the above example where the singlet neutrino masses are $\sim 10^{13}$ GeV,
the dimension-five operator, $\kappa_{ij} \ell_i H \ell_j H/M_\nu$ ($M_\nu$ stands for a scale
for the singlet neutrino masses, $M$ and $X$), 
is induced by loop correction.
The operator generates the light active neutrino masses when the Higgs field, $H$,
acquires an electroweak vev.
The loop-induced coupling matrix, $\kappa$, 
runs by RGE below the mass scale \cite{Babu:1993qv}.
As we have shown, we suppose 
that the extra gauge boson masses are the same scale of the singlet neutrinos
in order to realize the observed neutrino mass structure
(one of the gauge boson can be much heavier than the singlet neutrinos),
and therefore, the RGE effects by the extra gauge bosons are not significant.
The charged lepton Yukawa coupling for tau and muon
can additionally violate the $\mu$-$\tau$ symmetry. 
It is obvious that the flavor universal contributions in RGE does not change the neutrino mixing angles
but change the overall factors of neutrino masses.
The neutrino masses are surely chosen to realize the observed mass squared differences at low energy,
and we do not treat the overall factors in the discussions below.  
The Yukawa coupling, $y_\tau$, to the third generation can modify the mixings.
As a result, (neglecting to show the muon Yukawa contribution, which is much less than tau contribution)
the coupling of the dimension-five operators at the weak scale is
given as
\begin{equation}
\propto {\rm diag} (1, 1, z) \, \kappa_{ij} \, {\rm diag} (1, 1, z),
\end{equation}
where 
\begin{equation}
\frac{d}{d \ln \mu} \ln z = \frac{c}{16\pi^2} y_\tau^2,
\end{equation}
and 
\begin{equation}
z \simeq 1 - \frac{c}{16\pi^2} \int^{\ln \Lambda_\nu}_{\ln \Lambda_w} y_\tau^2 d (\ln \mu),
\end{equation}
where $\Lambda_w$ is a weak scale.
The coefficient $c$ is $-1/2$ for non-supersymmetric models, and $1$ for supersymmetric models.
In the model where the neutrino mass is induced by loop correction at a high scale,
we surely suppose non-supersymmetric,
and thus the RGE contribution too tiny to describe for $y_\tau \sim 0.01$.
Therefore, we can say that the neutrino mixing matrix which is constructed at a high scale
is hardly modified by RGE.
If one considers two Higgs doublet models (though there is no motivation to consider it in our model building), 
the tau Yukawa coupling can be larger
and $1-z$ can be a few percent, which can modify the mixing angles a little ($\sim 1^{\rm o}$).
%
The $\mu$-$\tau$ symmetric and breaking parameters
given in Eqs.(\ref{Mnu0}) and (\ref{Mnuprime}) (up to overall normalization)
are modified as
\begin{equation}
\left(
 \begin{array}{c}
   A \\ A^\prime 
 \end{array}
\right) 
\rightarrow
\frac12\left(
 \begin{array}{cc}
   1+z & 1-z \\
   1-z & 1+z 
 \end{array}
\right) 
\left(
 \begin{array}{c}
   A \\ A^\prime 
 \end{array}
\right),
\ \ 
\left(
 \begin{array}{c}
   B \\ B^\prime 
 \end{array}
\right) 
\rightarrow
\frac12\left(
 \begin{array}{cc}
   1+z^2 & 1-z^2 \\
   1-z^2 & 1+z^2 
 \end{array}
\right) 
\left(
 \begin{array}{c}
   B \\ B^\prime 
 \end{array}
\right),
\end{equation}
and $C \rightarrow z \, C$.
It is interesting to note that
the parametrization we use is useful by the redefinition above even if the 
$\mu$-$\tau$ breaking is generated by the RGE effects.
The essential points on the modifications of the mixing angles via RGE are well-investigated at the earlier stage,
 including the case of supersymmetric models
(see \cite{Haba:1999fk}, for example).
Only if the eigenvalues of the neutrinos are degenerate with aligned Majorana phases,
the mixing matrix is unstable under the RGE corrections.
The condition corresponds to $B \gg C$ in our parametrization,
and the reason of the destabilization can be easily understood from the diagonalization formula.
In our model where the neutrino mass is induced by the loop correction,
the eigenvalues are not degenerate (it is rather difficult to construct such mass degeneracy)
and the loop corrections do not destabilize the mixing matrix at the high scale.
In the case of the inverted hierarchy, two masses, $m_1$ and $m_2$, are degenerate
and there needs a fine-tuning basically.
However, the solar mixing angle is a free parameter 
in the models under the (approximate) $\mu$-$\tau$ symmetry,
and one should fit the mixing angles and the mass squared differences at low energy anyway.
We note that the mixing modifications by RGE starting from the $\mu$-$\tau$ reflection symmetry at a scale 
is studied in \cite{Nath:2018hjx}.

%
%

\section{Conclusion}

The separation of the $\mu$-$\tau$ symmetric and breaking terms is useful
to discuss the deviation from $\pi/4$ of the 2-3 neutrino mixing angle and
the Dirac CP phase $\delta$
as a parametrization of the neutrino mass matrix.
The description can be applied when one constructs any neutrino models
to reproduce the observed neutrino oscillation parameters.
We stress that the separation of the neutrino mass matrix is helpful to understand 
how the large CP violation with the nearly maximal 2-3 mixing is realized model-independently.
Since the observed 1-3 neutrino mixing is sizable
to keep the 2-3 mixing angle nearly maximal naively,
one may think that the $\mu$-$\tau$ exchange symmetry
is not a good frame to describe the observed pattern of the mixing angles.
However, we have shown that it leads a large CP violation in the neutrino sector.
In general, the CP phase is not necessarily equal to the Dirac CP phase $\delta$,
which can be measured by the neutrino oscillation experiments.
If the $\mu$-$\tau$ breaking parameters are fixed,
one can write a relation among the 2-3 and 1-3 mixings and the Dirac CP phase
such as given in Eqs.(\ref{eq8}), (\ref{eq9}), (\ref{eq10}).
It will be interesting to test such relations and to see if there is underlying symmetry
in the neutrino mass matrix.

We have built a model in which
the mixing of the gauge bosons which are even and odd under
the $\mu$-$\tau$ exchange symmetry
generates the $\mu$-$\tau$ breaking terms in the neutrino mass matrix
by loop effects.
The $\mu$-$\tau$ exchange symmetry is spontaneously broken 
in a hidden sector, and the extra gauge bosons 
are messengers to generate the $\mu$-$\tau$ breaking terms
in the neutrino mass matrix.
%
One can surely construct a model in which the scalar fields, which breaks
the $\mu$-$\tau$ exchange symmetry,
can couple to the neutrino sector via Yukawa-type interactions, 
but the pattern of the $\mu$-$\tau$ breaking is left to
the model-builder's discretion in such models.
In our model, on the other hand,
the pattern is determined
by the gauge boson masses and mixing
and the SM singlet heavy neutrino spectrum.
The pattern of the $\mu$-$\tau$ breaking in the neutrino mass
matrix can lead a large CP phase preferably.  
The extra gauge boson and SM singlet neutrinos 
can lie around TeV for Dirac neutrino mass to be around MeV.
The $B-L$ and $L_\mu - L_\tau$ gauge bosons can play
the $\mu$-$\tau$ even and odd interactions,
and then, their gauge couplings are constrained.

\section*{Acknowledgments}



 This work is partly supported by JSPS KAKENHI Grants No.17H01133 (T.F).

\appendix

\section*{Appendix A: Diagonalization of the neutrino matrix}

In this Appendix,
we show the (approximate) diagonalization matrices
of the neutrino mass matrix given in Section 2.

Before the description, we write a diagonalization matrix of 
$2\times 2$ matrix.
A $2\times 2$ symmetric matrix can be diagonalized as
\begin{equation}
\left(
 \begin{array}{cc}
   \cos \theta & - \sin\theta e^{i\phi} \\
   \sin \theta e^{-i \phi} & \cos\theta
 \end{array}
\right)
\left(
 \begin{array}{cc}
   a & b \\
   b & d
 \end{array}
\right)
\left(
 \begin{array}{cc}
   \cos \theta & \sin\theta e^{-i\phi} \\
   -\sin \theta e^{i \phi} & \cos\theta
 \end{array}
\right)
= 
\left(
 \begin{array}{cc}
   \lambda_1 & 0 \\
   0 & \lambda_2
 \end{array}
\right),
\end{equation}
where
\begin{equation}
\tan2 \theta = \frac{2b}{e^{i\phi} d - e^{-i\phi} a},
\qquad
\phi = {\rm arg} (ab^* + bd^*).
\end{equation}
The eigenvalues $\lambda_{1,2}$ (they are not eigenvalues unless the elements are real, though)
are
\begin{eqnarray}
\lambda_1 &=& a \cos^2 \theta - e^{i\phi} b \sin 2 \theta + e^{2i\phi} d \sin^2 \theta = 
e^{i\phi}\left(\frac{ae^{-i\phi}+ d e^{i\phi}}{2} - b \csc 2\theta \right), \\
\lambda_2 &=& d \cos^2 \theta + e^{-i\phi} b \sin 2 \theta + e^{-2i\phi} a \sin^2 \theta = e^{-i\phi} \left(\frac{a e^{-i\phi} + de^{i\phi}}{2}+ b \csc 2\theta \right).
\end{eqnarray}
%
We note that, for the solar neutrino mixing in the case of IH,
one needs,
\begin{equation}
|ae^{-i\phi}+ d e^{i\phi}| \ll 2 |b| \csc \theta_{12},
\end{equation}
to obtain $|\lambda_1| \simeq |\lambda_2|$.
If both $a$ and $d$ are much smaller than $b$, the solar mixing angle becomes nearly $\pi/4$,
and thus one needs a cancellation to obtain the mass degeneracy.

Now we move to the diagonalization of $3\times 3$ neutrino matrix, rotated by $U^{(23)}(-\pi/4,0)$,
\begin{equation}
\bar M_\nu =
\left(
 \begin{array}{ccc}
  D & \sqrt2 A & \sqrt2 A' \\
  \sqrt2 A & E & B' \\
  \sqrt2 A' & B' & F
 \end{array}
\right),
\end{equation}
which is $\bar M_\nu^0 + \bar M_\nu'$ given in Eqs.(\ref{Mnubar}) and (\ref{Mnuprime}).
The description of the diagonalization depends on the neutrino mass hierarchy.
We explain the diagonalization processes below for 3 cases:
general NH, IH cases, and NH for rank-2 matrix.

\medskip

$\bullet$ NH (general)

We assume $D, \sqrt2 A, \sqrt2 A', E, B' \sim \lambda F$, where $\lambda \sim 0.2$.
($D$ can be much smaller than the others).
The approximate PMNS matrix up to $O(\lambda^2)$
can be written as
\begin{equation}
U_{\rm PMNS} \simeq U^{(23)}(-\frac{\pi}4,0) U^{(23)}(\theta_{23}',\phi_1) U^{(13)}(\theta_{13}^0,\phi_2) U^{(12)}(\theta_{12}^0,\phi_3).
\label{PMNS-NH}
\end{equation}
The $\theta_{23}'$ angle rotation makes the (2,3) element of $\bar M_\nu$ to be zero,
and $\theta_{13}^0$ angle rotation makes the (1,3) element.
Then, a quantity $O(\lambda^2) F$ appears in (2,3) element, but we ignore it.
Finally, 1-2 block is diagonalized by $\theta_{12}^0$ rotation.
The diagonalization process of $\bar M_\nu$ can be illustrated by 
\begin{eqnarray}
\left(
 \begin{array}{ccc}
  \lambda {\tt X} & \lambda {\tt X} & \lambda {\tt X} \\
  \lambda {\tt X} & \lambda {\tt X} & \lambda {\tt X} \\
  \lambda {\tt X} & \lambda {\tt X} & 1  
 \end{array}
\right) \nonumber &\!\!\!&\!\!\!\!\!\!
\xrightarrow{U^{(23)}(\theta_{23}',\phi_1)} 
\left(
 \begin{array}{ccc}
  \lambda {\tt X} & \lambda {\tt X} & \lambda {\tt X} \\
  \lambda {\tt X} & \lambda {\tt X} & 0 \\
  \lambda {\tt X} & 0 & 1  
 \end{array}
\right) \\ &\!\!\!&\!\!\!\!\!\!
\xrightarrow{U^{(13)}(\theta_{13}^0,\phi_2)} 
\left(
 \begin{array}{ccc}
  \lambda {\tt X} & \lambda {\tt X} & 0 \\
  \lambda {\tt X} & \lambda {\tt X} & \lambda^2 {\tt X} \\
  0 & \lambda^2 {\tt X} & 1  
 \end{array}
\right)
\xrightarrow{U^{(12)}(\theta_{12}^0,\phi_3)} 
\left(
 \begin{array}{ccc}
  \lambda {\tt X} & 0 & \lambda^2 {\tt X} \\
  0 & \lambda {\tt X} & \lambda^2 {\tt X} \\
  \lambda^2 {\tt X} & \lambda^2 {\tt X} & 1  
 \end{array}
\right),
\end{eqnarray}
where $\tt X$'s denote arbitrary numbers.

One finds\footnote{
We remark that $\theta_{23}$ can be kept to be $\pi/4$ for $\phi_1 = \pm \pi/2$
even if $\theta_{23}' \sim 1$ ($B' \sim E \sim F$).
Since $\phi_1 = {\rm arg} (EB'^* + B' F^*) = {\rm arg} ({\rm Re}(BB'^*) + i \,{\rm Im}(C B'^*))$,
one finds $B'/B$ is pure imaginary, and $|B+B'| = |B-B'|$ for $\phi_1 = \pm \pi/2$.
Therefore, for $\phi_1 = \pm \pi/2$, it is more convenient to reparameterize $B'$ to be zero
by rephasing the fields before the 2-3 rotation by $U_0$,
though the physical Dirac CP phase surely does not depend on the diagonalization processes.
%
}
\begin{equation}
\tan^2\theta_{23} = \left| \frac{c_{23}' - e^{-i\phi_1} s_{23}'}{c_{23}' + e^{-i\phi_1} s_{23}'} \right|^2
=\frac{1-\sin2\theta'_{23} \cos\phi_1}{1+\sin2\theta'_{23} \cos\phi_1},
\end{equation}
and
\begin{equation}
\theta_{23} - \frac{\pi}4 \simeq - \frac12\sin2\theta_{23}' \cos\phi_1.
\end{equation}
Remarking that one can obtain
\begin{equation}
U^{(23)} (-\frac{\pi}4,0) U^{(23)} (\theta_{23}',\phi_1 ) = P U^{(23)} (\theta_{23}, \phi_1'),
\end{equation}
where $P$ is an unphysical diagonal phase matrix,
and $\phi_1' = \pi + 
 {\rm arg}(c_{23}' - e^{-i\phi_1} s_{23}')(c_{23}' + e^{-i\phi_1} s_{23}')= \pi+ \arctan (s_{23}'^2 \sin 2\phi_1/(1- 2 s_{23}'^2 \cos^2 \phi_1))$,
one finds that the Dirac CP phase is
\footnote{
%
We note
that
one can obtain an identity equation,
\begin{eqnarray}
&&U^{(23)}(\theta_{23}, \phi_1) U^{(13)}(\theta_{13}, \phi_2) U^{(12)} (\theta_{12},\phi_3) \\ \nonumber 
&=& {\rm diag}(e^{-i\phi_3},1,e^{i\phi_1}) U^{(23)}(\theta_{23},0) U^{(13)}(\theta_{13},\phi_2-\phi_1-\phi_3) U^{(12)} (\theta_{12},0){\rm diag}(e^{i\phi_3},1,e^{-i\phi_1}).
\end{eqnarray}
}
\begin{equation}
\delta \simeq \phi_2 - \phi_3 - \pi + O(s_{23}'^2).
\end{equation}
The 1-3 and 1-2 mixing angles corresponds to $\theta_{13}^0$ and $\theta_{12}^0$.
One can easily find that $A'$ leads to the 1-3 mixing and $B'$ provides $\delta \theta_{23}$.
If we suppose $A' = B'$, then $\theta_{13} = \theta_{23}'$ and $\phi_1=\phi_2$.
Thus, to make $\delta \theta_{23}$ in the allowed region, we need a large phase $\phi_1$.
However, the Dirac CP phase mixes with $\phi_3$ which is a phase to diagonalize the 1-2 block,
and the phase $\phi_1$ is not necessarily same as the Dirac CP phase, which is measured by the neutrino oscillations.

\medskip

$\bullet$ IH

We assume $\sqrt2 A', B', F \sim \lambda (\sqrt2 A, B,D)$, where $\lambda \sim 0.2$.
($F$ can be much smaller than the others).
The approximate PMNS matrix up to $O(\lambda^2)$
can be written as
\begin{equation}
U_{\rm PMNS} = U^{(23)} (-\pi/4,0) U^{(12)} ( \theta_{12}^0, \phi_1 ) U^{(13)} (\theta_{13}^\prime, \phi_2) U^{(23)} (\theta_{23}^\prime,\phi_3).
\label{PMNS-IH}
\end{equation}
By $U^{(23)}(-\pi/4,0)U^{(12)}(\theta_{12}^0,\phi_1)$,
the $\mu$-$\tau$ symmetric matrix is diagonalized as
$\tilde M_\nu^0 = {\rm diag} (m_1, m_2, F)$.
The $\mu$-$\tau$ breaking matrix is, then, in the basis, 
\begin{equation}
 \tilde M_\nu'
=
\left(
\begin{array}{ccc}
0 & 0 & \sqrt2 A'  c_{12}^0 -s_{12}^0 e^{i \phi_1} B'\\
0 & 0 &  \sqrt2 A'  s_{12}^0 e^{-i\phi_1} + c_{12}^0  B'\\
\sqrt2 A'  c_{12}^0 -s_{12}^0 e^{i \phi_1} B' & \sqrt2 A'  s_{12}^0 e^{-i\phi_1} + c_{12}^0  B'\ & 0
\end{array}
\right).
\label{UMU}
\end{equation}
We made 1-3 and 2-3 rotation further to eliminate the (1,3) and (2,3) elements.
Those two angles $\theta_{23}'$ and $\theta_{13}'$ are small, and we neglect the quantities 
in the off-diagonal elements
after the rotations.
The diagonalization process of $\bar M_\nu$ can be illustrated by 
\begin{eqnarray}
\left(
 \begin{array}{ccc}
   {\tt X} &  {\tt X} & \lambda {\tt X} \\
   {\tt X} &  {\tt X} & \lambda {\tt X} \\
  \lambda {\tt X} & \lambda {\tt X} & \lambda {\tt X}
 \end{array}
\right) \nonumber &\!\!\!\!&\!\!\!\!\!\!
\xrightarrow{U^{(12)}(\theta_{12}^0,\phi_1)} 
\left(
 \begin{array}{ccc}
  m_1 & 0& \lambda {\tt X} \\
  0& m_2 & \lambda {\tt X} \\
  \lambda {\tt X} & \lambda{\tt X} & \lambda {\tt X}
 \end{array}
\right)
\\
&\!\!\!&\!\!\!\!\!\!
\xrightarrow{U^{(13)}(\theta_{13}',\phi_2)} 
\left(
 \begin{array}{ccc}
  m_1' & \lambda^2 {\tt X} & 0 \\
  \lambda^2 {\tt X} & m_2 & \lambda {\tt X} \\
  0 & \lambda {\tt X} & \lambda {\tt X}
 \end{array}
\right)
\xrightarrow{U^{(23)}(\theta_{23}',\phi_3)} 
\left(
 \begin{array}{ccc}
  m_1' & \lambda^2 {\tt X} & \lambda^3 {\tt X} \\
  \lambda^2 {\tt X} & m_2' & 0 \\
  \lambda^3 {\tt X} & 0 & \lambda {\tt X}
 \end{array}
\right).
\end{eqnarray}
Because the diagonal elements are modified by the rotation,
primes ($^\prime$) are attached to them. 

We remark that the mixing angles $\theta_{ij}$ of a unitary matrix $U$ defined by the PDG convention
are obtained as
\begin{equation}
\tan\theta_{23} = \left|\frac{U_{23}}{U_{33}}\right|,
\qquad
\tan\theta_{12} = \left|\frac{U_{12}}{U_{11}}\right|,
\qquad
\sin\theta_{13} = \left| U_{13} \right|,
\end{equation}
and the CP phase $\delta$ can be extracted by using Jarlskog invariant
\begin{equation}
J = U_{12} U_{23} U_{13}^* U_{22}^* .
\end{equation}
Expanding Eq.(\ref{PMNS-NH}), we obtain
\begin{eqnarray}
\tan\theta_{23} &=& 
\left| \frac{c_{13}'+ e^{i(\phi_1-\phi_2)} s_{12}^0 s_{13}' - e^{-i\phi_3} c_{12}^0  t_{23}' }{c_{13}'- e^{i(\phi_1-\phi_2)} s_{12}^0 s_{13}' + e^{-i\phi_3} c_{12}^0  t_{23}' } \right|,
\\
\tan\theta_{12} &=& \left| \frac{c_{23}'}{c_{13}'} t_{12}^0 - e^{i (\phi_1-\phi_2+\phi_3)} s_{23}' t_{13}'     \right| =  |\tan\theta_{12}^0 + O(\lambda^2)|, \\
\sin\theta_{13} &=& \left|s_{12}^0 s_{23}^\prime e^{-i\phi_3} + c_{12}^0 c_{23}' s_{13}^\prime e^{i (\phi_1-\phi_2)} \right|,  \\
J &=& - \frac14 \sin2\theta_{12}^0 c_{13}' c_{23}'^3 (e^{i\phi_3}  s_{12}^0 s_{23}^\prime +
e^{i(\phi_2 - \phi_1)}  c_{12}^0  c_{23}' s_{13}^\prime) + O(\lambda^2). \label{jarlskog}
\end{eqnarray}
The approximate expressions are rewritten as\footnote{
Under the PDG convention, one obtains
\begin{equation}
\delta = {\rm arg} (J+ s_{12}^2 s_{23}^2 s_{13}^2 c_{13}^2).
\end{equation}
The $\sim \lambda^2$ terms of the expansion in Eq.(\ref{jarlskog}) 
corresponds to the $s_{12}^2 s_{23}^2 s_{13}^2 c_{13}^2$ term,
and thus, one finds 
$\delta \simeq {\rm arg} (-\bar\Theta_{13})$.
}
\begin{equation}
\tan\theta_{23} 
\simeq \left|
\frac{1+\Delta}{1-\Delta}
\right|, 
\qquad
\sin\theta_{13} \simeq |\Theta_{13}|, \qquad
\delta \simeq \pi-{\rm arg} \Theta_{13},
\end{equation}
\begin{equation}
\Delta \equiv ( -c_{12}^0 s_{23}' e^{-i\phi_3} + s_{12}^0 s_{13}' e^{i \phi_{12}}),
\qquad
\Theta_{13} \equiv s_{12} s_{23}' e^{-i\phi_3} + c_{12} s_{13}' e^{i\phi_{12}},
\end{equation}
%
where
$\phi_{12} = \phi_1-\phi_2$.
We note that one obtains by Taylor expansion as
\begin{equation}
\arctan \left| \frac{1+\Delta}{1-\Delta} \right|  - \frac{\pi}4 = {\rm Re}\, \Delta + O(\Delta^3),
\end{equation}
and $O(\Delta^2)$ is absent, and thus, it gives a good approximation up to $O(\lambda^3)$.
Defining 
\begin{equation}
\frac{s_{13}'}{s_{23}'} = \tan\theta'.
\end{equation}
we express 
\begin{equation}
\theta_{23} - \frac{\pi}4 \simeq  {\rm Re} \left( \frac{-c_{12} c' e^{-i\phi_3} + s_{12} s' e^{i\phi_{12}}}{s_{12} c' e^{-i\phi_3} + c_{12} s' e^{i\phi_{12}}} \Theta_{13} \right).
\end{equation}
Calculating the real part of the expression, one obtains
\begin{eqnarray} \label{relation}
\theta_{23} - \frac{\pi}4 &\simeq& \frac{s_{12} c_{12} (c'^2-s'^2) \cos\delta
- c's' (s_{12}^2 \cos(\delta - \phi_{123}) -c_{12}^2 \cos(\delta + \phi_{123}))
}{s_{12}^2 c'^2 + c_{12}^2 s'^2 + 2 s_{12} c_{12} c's' \cos\phi_{123} } \theta_{13}\\
&=&\frac12
\frac{\sin2\theta_{12} \cos2\theta' \cos\delta+ \sin2\theta' (\cos2\theta_{12} \cos\delta\cos\phi_{123} - \sin\delta \sin\phi_{123})}{s_{12}^2 c'^2 + c_{12}^2 s'^2 + 2 s_{12} c_{12} c's' \cos\phi_{123} } \theta_{13}, \nonumber
\end{eqnarray}
where $\phi_{123} = \phi_{12}+\phi_3$.
If one tune as
\begin{equation}
\sin2\theta_{12}\cot2\theta' \cos\delta + \cos 2\theta_{12} \cos\phi_{123}\cos\delta = \sin\delta\sin\phi_{123},
\end{equation}
$\theta_{23}$ can be kept to be $\pi/4$ irrespective of the CP phases.
As a consequence, the CP phase cannot be predicted in general.
However, if one wants to avoid such a cancellation between $A'$ and $B'$ contributions,
the choice of $\cos\delta \simeq 0$ and $\sin\phi_{123}\simeq 0$ can be a simple solution
to avoid a large deviation from $\pi/4$.
If one of $A'$ and $B'$ is zero as a special point, such cancellation is avoided obviously,
and a relation among $\theta_{13},\delta \theta_{23}$ and $\delta$ can be obtained.
Let us calculate the relation in such cases.
For $A' = 0$, one obtains $\tan\theta' = \tan\theta_{12}$.
For $B' = 0$, one obtains $\tan\theta' = -\cot \theta_{12}$.
Remarking $m_1 : m_2 \simeq e^{i\phi_1} : - e^{-i\phi_1}$ in the convention we use,
one finds $\phi_{123} = 0$.
As a result, we obtain Eqs.(\ref{eq9}) and (\ref{eq10}), which can be verified by numerical calculations.

\medskip

$\bullet$ NH (rank-2 matrix)

The reason why the PMNS matrix for NH in Eq.(\ref{PMNS-NH})
is different from
the one in IH given in Eq.(\ref{PMNS-IH})
is that
 $O(\lambda)$ mixing remains in 1-2 block in general
 and one needs additional 1-2 rotation
if the diagonalization is executed by Eq.(\ref{PMNS-IH}):
\begin{eqnarray}
\left(
 \begin{array}{ccc}
  \lambda {\tt X} & \lambda {\tt X} & \lambda {\tt X} \\
  \lambda {\tt X} & \lambda {\tt X} & \lambda {\tt X} \\
  \lambda {\tt X} & \lambda {\tt X} & 1  
 \end{array}
\right) \nonumber &\!\!\!\!&\!\!\!\!\!\!
\xrightarrow{U^{(12)}(\theta_{12}^0,\phi_1)} 
\left(
 \begin{array}{ccc}
  \lambda {\tt X} & 0 & \lambda {\tt X} \\
  0 & \lambda {\tt X} & \lambda {\tt X} \\
  \lambda {\tt X} & \lambda {\tt X} & 1  
 \end{array}
\right) \\ &\!\!\!&\!\!\!\!\!\!
\xrightarrow{U^{(13)}(\theta_{13}',\phi_2)} 
\left(
 \begin{array}{ccc}
  \lambda {\tt X} & \lambda^2 {\tt X} & 0 \\
  \lambda^2 {\tt X} & \lambda {\tt X} & \lambda {\tt X} \\
  0 & \lambda {\tt X} & 1  
 \end{array}
\right)
\xrightarrow{U^{(23)}(\theta_{23}',\phi_3)} 
\left(
 \begin{array}{ccc}
  \lambda {\tt X} & \lambda^2 {\tt X} & \lambda^3 {\tt X} \\
  \lambda^2 {\tt X} & \lambda {\tt X} & 0 \\
  \lambda^3 {\tt X} & 0 & 1  
 \end{array}
\right).
\end{eqnarray}
If the (1,3) element of $\tilde M_\nu'$, $\sqrt2 A' c_{12}^0 - s_{12}^0 e^{i\phi_1 B'}$, given in Eq.(\ref{UMU})
is small, such additional 1-2 rotation is not needed, 
and thus, the PMNS matrix for NH can be also given as in Eq.(\ref{PMNS-IH})\footnote{
The PMNS matrix in NH case can be also given by Eq.(\ref{PMNS-IH})
if the (2,3) element of $\tilde M_\nu$ is small.
In this case, one finds $s_{23}'=0$ ($c'=0,s'=1)$ in Eq.(\ref{relation}),
and the relation is given as $\theta_{23} -\pi/4 \simeq - \tan\theta_{12} \sin\theta_{13} \cos\delta$.
}.
One can find that the condition $\sqrt2 A' c_{12}^0 - s_{12}^0 e^{i\phi_1 B'} = 0$
is satisfied if the $\mu$-$\tau$ symmetric matrix and the total neutrino mass matrix are rank 2.
%
Then, 
the relation is given by Eq.(\ref{relation})
with $s_{13}' = 0$ $(\tan\theta'=0)$.
The diagonalization process of $\bar M_\nu$ for the rank-2 case can be illustrated by 
\begin{equation}
\left(
 \begin{array}{ccc}
  \lambda a^2 & \lambda a b & \lambda a c \\
  \lambda a b & \lambda b^2 & \lambda b c \\
  \lambda a c & \lambda b c & d 
 \end{array}
\right)
\xrightarrow{U^{(12)}(\theta_{12}^0,\phi_1)} 
\left(
 \begin{array}{ccc}
  0  & 0 & 0 \\
  0 & \lambda X^2 & \lambda X c \\
  0 & \lambda X c & d
 \end{array}
\right)
\xrightarrow{U^{(23)}(\theta_{23}',\phi_3)} 
\left(
 \begin{array}{ccc}
  0 & 0 & 0 \\
  0 & \lambda X'^2 & 0 \\
  0 & 0 & d'
 \end{array}
\right),
\end{equation}
where $X=\sqrt{a^2+b^2}$,
and primes attached to the diagonal elements
are because of the modification by the last 2-3 rotation.
As a result, we find that the relation for rank-2 matrix in NH is given as in Eq.(\ref{eq8}).
One can verify the relation by numerical diagonalization of the neutrino mass matrix.

We comment on the case of general rank-2 neutrino matrix.
The rank-2 matrix can be given by two row vectors $x_1$ and $x_2$ as
$M_\nu = x_1^T x_1 + x_2^T x_2$. Without loss of generality (up to unphysical redefinitions),
one can parametrize as $x_1 = (e, -d, d)$ and $x_2 = (x,y,z)$, for $e, x, y, z \sim \lambda d$.
This can be understood as that the rank-2 symmetric matrix has 5 complex degrees of freedom.
One can find that $e\neq 0$ and $y-z \neq 0$ break the $\mu$-$\tau$ exchange symmetry.
The PMNS matrix for the rank-2 matrix can be written as
\begin{equation}
U^{\rm PMNS} = U^{(23)}(-\frac{\pi}4,0) U^{(13)}(\theta_{13}',\phi_1) U^{(12)}(\theta_{12}^0,\phi_2) U^{(23)} (\theta_{23}',\phi_3).
\end{equation}
By the $U^{(23)}(-\frac{\pi}4,0) U^{(13)}(\theta_{13}',\phi_1)$ rotation,
$x_1 \to x_1' \propto (0,0,1)$.
The $U^{(12)}(\theta_{12}^0,\phi_2)$ rotation makes the 1st element of $x_2$ to be zero.
Finally, $U^{(23)} (\theta_{23}',\phi_3)$ diagonalize the remained $2\times 2$ matrix.
Expanding the expression similarly to the previous case,
one finds
\begin{equation}
\theta_{23} - \frac{\pi}4 \simeq \theta_{13} \frac{c_{12} s_{23}' (s_{12} s_{23}' \cos\delta + s_{13}' \cos(\phi_{123} - \delta))}{s_{13}'^2+ s_{12}^2 s_{23}'^2 + s_{13}' s_{12} s_{23}' \cos\phi_{123}}.
\end{equation}
One can find that $\theta_{13}' = 0$ if $e=0$, and then, Eq.(\ref{eq8}) holds.

\section*{Appendix B: Comparison with the $\mu$-$\tau$ reflection symmetry }

We comment on the $\mu$-$\tau$ reflection symmetry 
considered in 
\cite{Harrison:2002et}.
Lagrangian is assumed to be invariant
under $\nu_e \leftrightarrow \nu_e^*$, $\nu_\mu \leftrightarrow \nu_\tau^*$,
and thus,
the neutrino mass matrix is given as
\begin{equation}
M_\nu = \left(
\begin{array}{ccc}
 W & X & X^* \\
 X & Y & Z \\
 X^* & Z & Y^*
\end{array}
\right),
\end{equation}
where $W$ and $Z$ are real.
By the field redefinitions $\nu_\mu \to \nu_\mu \eta$,
$\nu_\tau \to \nu_\tau \eta^*$,
$\eta^2 = e^{-i\,{{\rm arg}\, Y}}$,
one obtains $Y \to |Y|$, $X \to X \eta$.
Therefore, without loss of generality, $Y$ can be also considered to be real
and only $X$ is complex.
Comparing with Eqs.(\ref{Mnubar}) and (\ref{Mnuprime}),
one finds
\begin{equation}
U_0^T P_\eta M_\nu P_\eta U_0 = 
\left(
 \begin{array}{ccc}
  W & \sqrt2\, {\rm Re} (X\eta) & -i \sqrt2\, {\rm Im} (X\eta) \\
  \sqrt2\, {\rm Re} (X\eta) & |Y|+Z & 0 \\
  -i \sqrt2\, {\rm Im} (X\eta) & 0 & |Y|- Z
 \end{array}
\right),
\ 
P_\eta = {\rm diag} ( 1, \eta, \eta^*),
\end{equation}
and $A = {\rm Re} (X\eta)$, $A' = - i \,{\rm Im} (X \eta)$, $B' =0$.
Because $A'$ is pure imaginary and the other elements are real,
one easily finds $\delta = \pm \pi/2$ and $\theta_{23} = \pi/4$.
The $\mu$-$\tau$ reflection symmetry is a kind of ``CP" transformation property
to make $\delta = \pm \pi/2$.
On the other hand,
we do not assume any CP property under the $\mu$-$\tau$ exchange symmetry.
Even without the CP property, we find that a large CP violation is preferable 
in the neutrino mass matrix with the $\mu$-$\tau$ exchange symmetry breaking
to keep the 2-3 mixing angle to be nearly maximal, as explained in the text.
In fact, the configuration of $\mu$-$\tau$ reflection symmetry
is a special choice of the $\mu$-$\tau$ exchange symmetry breaking.
One can easily construct a model with $\mu$-$\tau$ reflection symmetry,
if a scalar field, which breaks the $\mu$-$\tau$ exchange symmetry,
also breaks CP symmetry spontaneously.

\end{document}